\documentclass{elsart}

 \usepackage{graphics}
 \usepackage{graphicx}
 \usepackage{epsfig}

\usepackage{amssymb}

\begin{document}

\begin{frontmatter}



\title{Magnetic dipole probes of the $sd$ and $pf$ shell 
        crossing in the $^{36,38}$Ar isotopes\thanksref{*}}
\thanks[*]{This work has been supported through the SFB634 of the Deutsche Forschungsgemeinschaft.}

\author[a]{A.~F.~Lisetskiy}\ead{olisetsk@theory.gsi.de}
\author[b]{E. Caurier}
\author[a,c]{K. Langanke}
\author[a]{G. Mart\'inez-Pinedo}
\author[c]{P. von Neumann-Cosel}
\author[b]{F. Nowacki}
\author[c]{A. Richter}

\address[a]{GSI, Planckstr.1, D-64291, Darmstadt, Germany}
\address[b]{Institut de Recherches Subatomiques, Universit\'e Louis Pasteur, 67037 Strasbourg, France}
\address[c]{Institut f\"ur Kernphysik, Technische Universit\"at Darmstadt, D-64289, Darmstadt, Germany}

\begin{abstract}
We have calculated the M1 strength distributions in the
$^{36,38}$Ar isotopes within large-scale shell model studies
which consider valence nucleons in the $sd$ and $pf$ shells.
While the M1 strength in $^{36}$Ar is well reproduced within
the $sd$ shell, the experimentally observed strong fragmentation
of the M1 strength in $^{38}$Ar requires configuration mixing 
between the $sd$ and the $pf$ shells adding to our understanding of
correlations across the N=20 shell gap.
\end{abstract}

\end{frontmatter}

\section{Introduction}
\label{1}

The shell model concepts of {\em closures}, {\em magic numbers} and
{\em inert cores}
leads to a reasonably good description of nuclear phenomena in
light as well as in heavy
nuclei.
However, recent experimental findings show that shell closures might get eroded
in very neutron-rich nuclei and that the inter-shell correlations
among nucleons play a decisive role for the structure of these nuclei.
A famous example is the nucleus $^{32}$Mg which is strongly deformed
in the ground state, despite its neutron number $N=20$, e.g. \cite{Motobayashi}.
\begin{figure}
\includegraphics[scale=0.35]{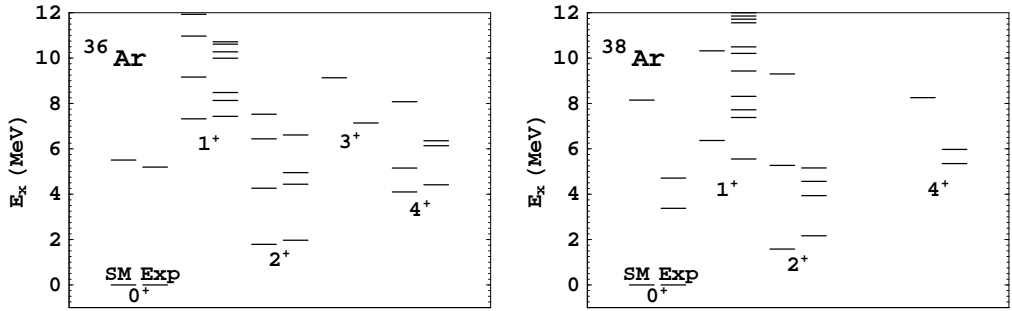}
\caption{Calculated and experimental spectra for $^{36}$Ar (left) 
and $^{38}$Ar (right). Excitation energy $E_x$ is plotted along the y-axis.
 The calculations have been performed in the
$sd$ shell using the USD interaction \cite{usd}. The model space allows only
 the formation of positive parity states. }
\label{fig1}
\end{figure} 
Thus, shell closures are not 'rigid' and there have been for a
long time indications for cross-shell correlations even in nuclei close to
classical double-magic nuclei like $^{40}$Ca. For this nucleus,
these include the observation of a strong M1 transition \cite{Gross79} and of a 
non-vanishing Gamow-Teller (GT) strength built on the $^{40}$Ca ground state \cite{pn1,pn2,np}. 
On the theory side, recent studies \cite{Cau01,Schi03} show that there are 
very large admixtures of $sd$ configurations in low-energy states of Ca
isotopes with $A\ge40$, although these nuclei are usually considered as 
"good" $pf$ shell nuclei.
\begin{figure}
\includegraphics[scale=0.34]{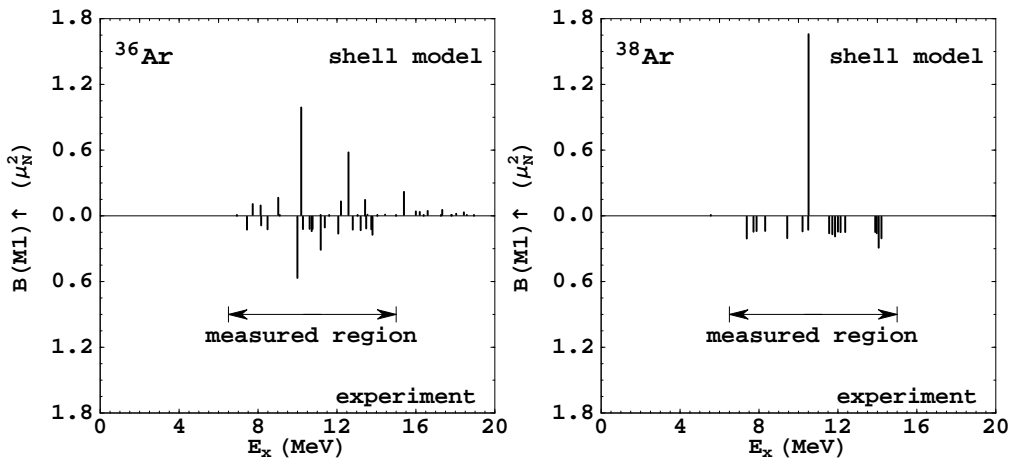}
\caption{Calculated and experimental M1 $0^+_1 \rightarrow 1^+_f$ strength 
distributions for $^{36}$Ar (left) and $^{38}$Ar (right). The calculations have been 
performed in the $sd$-shell using the USD interaction \cite{usd}. The 
spin part of the M1 operator has been quenched by a factor of 0.75
which is in agreement with the one 
deduced for other nuclei in this mass range \cite{PNC98}.
The same quenching of the spin part of the M1 operator
has been used throughout this paper.
Experimental error bars, which are of order of 10$\%$ for all 
measured states, are not shown here and in all further figures.}
\label{fig2}
\end{figure} 
Another indication for the erosion of the $sd$--$pf$ closure near $A=40$ are 
the data for the M1 strength in $^{36}$Ar and  $^{38}$Ar \cite{Fol94}.
While experimentally the M1 strength distribution for both nuclei is rather 
fragmented, this feature is reproduced by $sd$ shell model calculations for 
$^{36}$Ar, while these studies predict only one strong M1 transition for $^{38}$Ar,
in clear contradiction to the data \cite{Fol94}.
Obviously a single $sd$-shell model space, allowing only for two proton holes 
with respect to a $^{40}$Ca core, besides a closed neutron shell, is too 
limited for a description of the M1 data in $^{38}$Ar.

The fact that cross-shell excitations are quite important is revealed by a 
detailed comparison of the experimental and calculated spectrum at low energies.
The shell model calculations have been performed within the $sd$ shell using 
the standard USD interaction \cite{usd}. 
The model space does not allow the formation of negative parity states,
but the calculation also misses a few of the low-lying $0^+$ and $2^+$
states in $^{38}$Ar, while the agreement between theory and experiment
is quite sufficient for the positive parity states in $^{36}$Ar (see Fig.\ref{fig1}).
In Fig.\ref{fig2} we compare the experimental M1 strength distribution
with the shell model results, which are similar to those published earlier in
 \cite{Fol94}. The shell model code ANTOINE \cite{code} has been used for all 
calculations presented in this paper.

The detailed comparison of the $sd$-shell model results with the data shows
that the M1 data for the two argon isotopes $^{36}$Ar and $^{38}$Ar
present a challenge to theory: correlations between $sd$ and $pf$ shells should
not affect the major features of the M1 strength distribution for $^{36}$Ar,
but must be essential for the $^{38}$Ar M1 data, e.g. if only two neutrons
are added to the nucleus. 

The present paper is an attempt to reproduce this striking behavior
on the basis of large-scale shell model calculations which
combine both $sd$ and $pf$ shells. We have followed two routes.
At first, we have performed studies within the complete $sdp\!f$ shell.
However, due to computational limitations these calculations had to be
restricted to 2-particle 2-hole (2p2h) and 4p4h excitations from the $sd$- to the $pf$-shell.
These studies were supplemented by full diagonalization calculations within
the $d_{3/2}s_{1/2}f_{7/2}p_{3/2}$ model space (denoted as $sdp\!f_0$ space below).
Further additional 1p1h coupling between the 
$d_{5/2}$, $f_{5/2}$, $p_{1/2}$ orbitals and the $sdp\!f_0$ space 
has been taken into account  -- this extended space is refered to as 
$sdp\!f_1$ below.  As we will show in the following
sections, the calculations in the $sdp\!f_1$ model space indeed result in 
fragmentation of the M1 strength for both argon isotopes and also reproduce their
low-energy excitation spectrum.

\section{Truncated shell model studies in the complete $sdp\!f$ shells}

There have been several attempts to derive a suitable effective interaction for 
the $sdp\!f$ model space (see, for example \cite{War86,Ret97,Num01} and references therein).
For our study we adopt the one of Ref. \cite{Num01}. This interaction contains three 
parts: the USD interaction for the  $sd$-shell \cite{usd}, the KB' matrix elements for the
$pf$ shell \cite{KB} and the G-matrix of Kahana, Lee and Scott for
the cross-shell matrix elements \cite{GKahana}.
The  $f_{7/2}d_{3/2}$, $f_{7/2}s_{1/2}$, $p_{3/2}s_{1/2}$ and  $p_{3/2}d_{3/2}$
monopole terms have been adjusted to some recent experimental data \cite{Num01}.
We note hat there is a problem with spurious center-of-mass (COM) excitations for the 
chosen configuration space. To treat this problem we have added the tenfold of the 
center-of-mass Hamiltonian to the original Hamiltonian \cite{Num01}.
This procedure pushes states with predominantly center-of-mass excitations
up in energy and practically eliminates the spurious excitations in the low-energy states 
of interest here.

Unfortunately full diagonalization studies are yet not possible in the
complete $sdp\!f$ model space due to computational limitations.
However, it is feasible to perform calculations for $^{38}$Ar
in which 2 particles or even 4 particles are promoted from the $sd$ shell
to the $pf$ shell. We will refer to these studies as
2p2h and 4p4h calculations, respectively. The dimension of the 4p4h model space 
is 227 625 436 for $M=0^+$ states in the m-coupling scheme. 
For $^{36}$Ar we have performed 2p2h calculations, corresponding to 3 148 356 $M=0^+$ 
states.

These studies, however, showed no improvement for the $^{38}$Ar spectrum
compared to the pure $sd$-shell results. 
In fact the first excited $0^+$ state
lies at an excitation energy above 7 MeV, with very little
mixing to the ground state which had basically an $sd$ shell configuration.
Consequently no fragmentation has been found in our calculation of the M1 strength 
distributions for $^{38}$Ar.  We also have performed No-Core 
Shell Model \cite{NCSM1,NCSM2} calculations
in a $4\hbar\omega$ model space using the UCOM interaction \cite{ucom} obtaining similar results.
\begin{figure}
\includegraphics[scale=0.56]{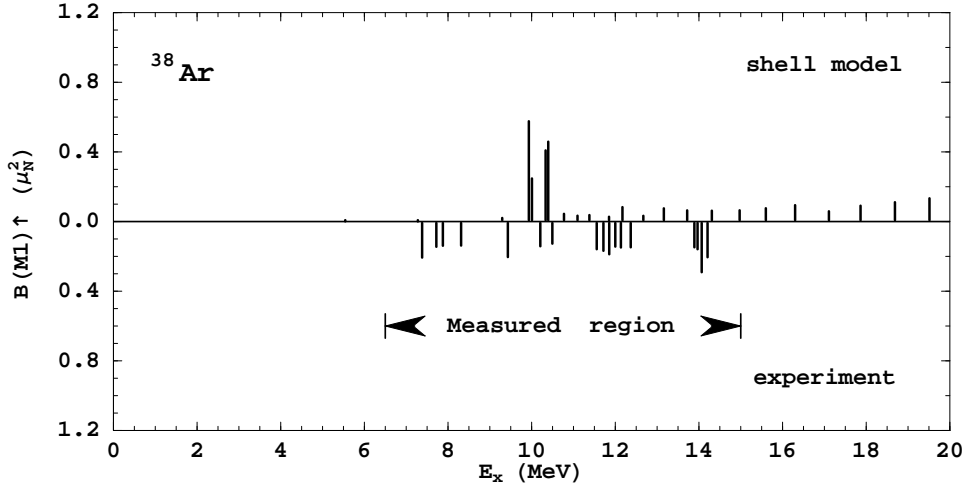}
\caption{Calculated (upper panel) and experimental (lower panel)   
M1 strength distributions in $^{38}$Ar in the full $sdp\!f$ space with 
4p4h cross-shell excitations.  
The calculations have been performed using a monopole-modified interaction
as described in the text. 
}
\label{fig3}
\end{figure} 

The mixing of the $sd$ and $pf$ shell configurations can be enhanced by slight monopole
modifications in the interaction. This strategy has been applied in large-scale shell model 
calculations of $^{36}$Ar and leads to a quite satisfying description of the
strongly deformed 4p4h band and its coupling to the ground state band \cite{Poves}.
Following this reference we have lowered the  energy of the $n$p$n$h configurations 
with $n=4$ and $n=6$ by suitable monopole shifts, in this way enforcing $pf$-shell configurations
to mix into the ground states. As it is expected the mixing results in fragmentation of 
the M1 strength distribution. This can be seen in Fig.\ref{fig3} which shows the calculated M1 
strength for $^{38}$Ar obtained in a 4p4h calculation and employing Lanczos diagonalization with 
50 iterations. Indeed the strong M1 transition observed in the pure $sd$-shell calculation 
is now split into several components and some M1 strength is shifted to higher
excitation energies. Nevertheless, the agreement with data is only marginal. 
In particular, the data show already some M1 strength around 8 MeV, which is missing in the 
calculation. This is possibly improved if $n$p$n$h configurations with $n>4$ were included -- previous 
shell model studies \cite{Cau01,Schi03,Ret97,Num01} indicated that energies of the states
in nuclei around $^{40}$Ca are converging rather slowly when  $n$p$n$h 
cross-shell excitations with larger $n$ are gradually added.  While such converged calculations 
are yet impossible in the complete $sdp\!f$ shell, they are, however, feasible in the 
$sdp\!f_0$ and $sdp\!f_1$ model spaces (see text above) to which we turn in the next section.

\section{Large-scale shell-model calculations in the truncated $sdp\!f$ model space} 

\begin{figure}[t]
\begin{center}
\includegraphics[scale=0.35]{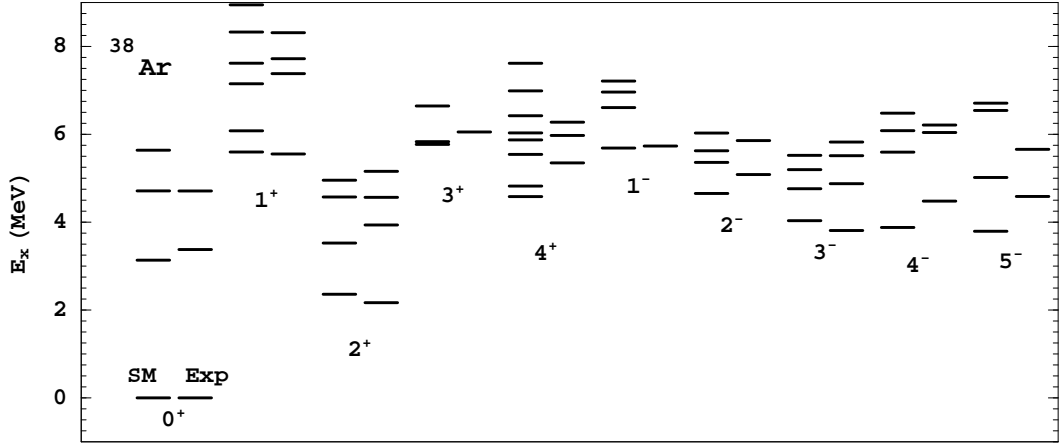}
\end{center}
\caption{Calculated and experimental spectra for $^{38}$Ar within the full 
 $sdp\!f_1$ space.}
\label{fig4}
\end{figure}

\begin{figure}[b]
\begin{center}
\includegraphics[scale=0.35]{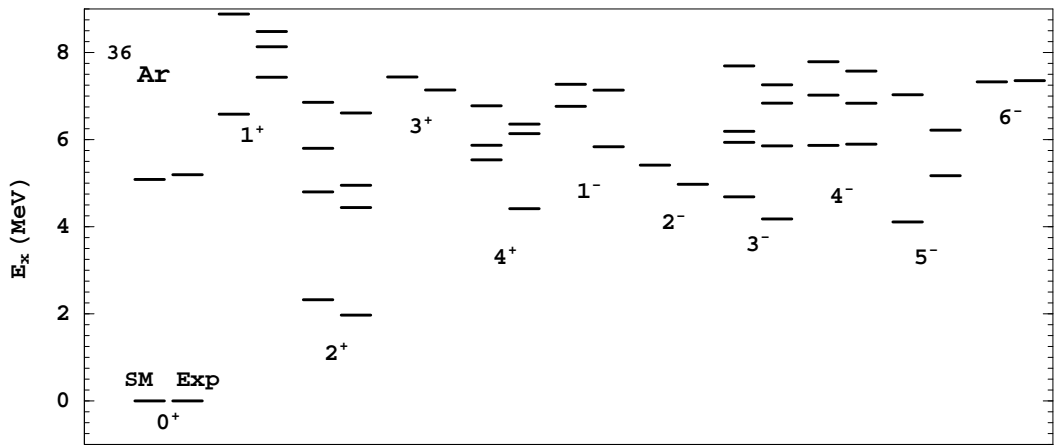}
\end{center}
\caption{Calculated and experimental spectra for $^{36}$Ar within the full 
 $sdp\!f_1$ space.}
\label{fig5}
\end{figure}

We have performed diagonalization calculations in the  $sdp\!f_1$ model space 
which includes the spin-orbit partners (missing in the $sdp\!f_0$ space) approximately
 by allowing 1p1h excitations from the $d_{5/2}$ orbital to the rest of the 
$sd$ shell, and from the $f_{7/2}p_{3/2}$ space to the $f_{5/2}p_{1/2}$ sub-space,
respectively. In other words we put 1p1h predominantly spin-flip excitations 
on top of the complex $n$p$n$h $(s_{1/2} d_{3/2})^{A-28-n}(f_{7/2} p_{3/2})^n$ cross-shell 
configurations obtained within the $sdp\!f_0$ space. This extension increases the dimension of the
 model space from 2.3 million ($sdp\!f_0$)  to nearly 75 million ($sdp\!f_1$) for $^{38}$Ar.
For the calculations we have used the same effective interaction as in \cite{Cau01}.
The $sdp\!f_0$ model space allows up to 8 or 10 particles to be promoted from the $sd$-shell
 to the $pf$-shell for $^{36}$Ar and $^{38}$Ar, respectively. It turns out these cross-shell 
excitations are sufficient to achieve a good  agreement between our calculations and the 
experimental $^{38}$Ar spectrum. As one can observe in Fig. \ref{fig4},
 the calculation now 
yields the low-lying $0^+$ and $2^+$ excitations which were missing in 
the $sd$-model space (see Fig. \ref{fig1}). The calculated $^{36}$Ar spectrum in the 
$sdp\!f_1$ model space (see Fig. \ref{fig5}) agrees similarly well with the data as the pure $sd$-shell 
result (Fig. \ref{fig1}) for the positive parity states. For both argon isotopes the calculated 
negative parity spectrum, missing in the pure $sd$-shell studies,  agrees quite well
with the observed one.
\begin{table*} 
\caption{The structure of the ground state and
of two final states with strong M1 transitions
(the $1^+,T=1$ at 10.14 MeV and 
the $1^+,T=1$ at 14.77 MeV states) for $^{36}$Ar in the full $sdp\!f_1$ space. 
The coefficient $A_{n}(n_{df})$ gives the contribution 
of the component with $n$ particles promoted across the $sd$--$pf$ shell gap. The 
 $n_{df}$ quantity indicates whether 1p1h spin-flip components are present
 ($n_{df}=1$) or not ($n_{df}=0$). Only components with weights larger than 
0.01 are shown. } 
\vspace{0.2cm}
\label{tab1} 
\begin{center} 
\begin{tabular}{c|ccc} 
 \hline 
\multicolumn{4}{c}{ $J_i^\pi = 0^+_1,T=0$  } \\
\hline
       $n$      &    0   &  2    &   4       \\
\hline 
 $A_{n}(0)$   &   0.47 & 0.30  & 0.05    \\  
 $A_{n}(1)$   &   0.01 & 0.11  & 0.04    \\  
\hline 
\multicolumn{4}{c}{ $J_i^\pi = 1^+,T=1$ ($E_x=10.14$ MeV) } \\
\hline
       $n$        &   0  &  2     &  4      \\ 
 $A_{n}(0)$   &   0.14 & 0.51   & 0.08    \\  
 $A_{n}(1)$   &   0.03 & 0.13   & 0.09    \\  
\hline 
\multicolumn{4}{c}{ $J_i^\pi = 1^+,T=1$ ($E_x=14.77$ MeV) } \\
\hline
       $n$        &   0    & 2      &  4       \\ 
 $A_{n}(0)$     &   0.02 & 0.38   & 0.15    \\  
 $A_{n}(1)$     &  0.07  & 0.23   & 0.14     \\  
\hline 
\end{tabular} 
\end{center} 
\end{table*} 

It is interesting to inspect more closely the role of $n$p$n$h cross-shell 
and 1p1h spin-flip 
excitations in the $sdp\!f_1$ space for the
$^{36}$Ar and $^{38}$Ar ground states. To quantify the importance of $n$p$n$h 
cross-shell
excitations we have calculated the weights $A_n(n_{df}=n_d+n_f)$ of the
$$(d_{5/2})^{6-n_{d}}(s_{1/2} d_{3/2})^{A-28-n+n_{d}}(f_{7/2} p_{3/2})^{n-n_f}(f_{5/2} p_{1/2})^{n_f}$$
configurations, where $n_{df}=n_d+n_f $ define the total number
of excitations from the $d_{5/2}$ orbital to the rest of the $sd$-shell
and from the ($f_{7/2}p_{3/2}$) subspace to the 
respective spin-orbit partners. Our calculations are restricted to $n_{df}=1$.
Thus,  
with $A_n(n_{df}=1)$ we have identified the components of the wave functions
which arise from 1p1h couplings to the spin-orbit partners within the
$sd$-shell ($n_d =1,n_f=0$) or $pf$-shell ($n_d =0,n_f=1$), respectively.
 The dominant configurations of the ground state wave
 functions are given in Tables \ref{tab1} and \ref{tab2} for $^{36}$Ar and $^{38}$Ar, respectively.
In the $sdp\!f_1$ model space, approximately half of the $^{36}$Ar ground state
corresponds to  pure $sd$-shell configurations, where the 1p1h
excitations of the $d_{5/2}$ orbit play a tiny role ($1\%$).

Cross-shell 2p2h components contribute about $41\%$ to the
$^{36}$Ar ground state, where the couplings to the spin-orbit
partners are now sizable ($11\%$). Higher-order cross-shell components
with $n>2$ are relatively unimportant (less than $10\%$).
These components, however, are larger in the $^{38}$Ar ground state.
\begin{table*} 
\caption{The structure of the ground state
($0^+,T=1$) and of two final states with sizable M1 transitions
(the $1^+,T=1$ at 12.96 MeV and 
the $1^+,T=2$ at 18.22 MeV states) for $^{38}$Ar in the full $sdp\!f_1$ space. 
The coefficient $A_{n}(n_{df})$ gives the contribution 
of the component with $n$ particles promoted across the $sd$--$pf$ shell gap. The 
 $n_{df}$ quantity indicates whether 1p1h spin-flip components are present
 ($n_{df}=1$) or not ($n_{df}=0$). Only components with weights larger than 
0.01 are shown.} 
\vspace{0.2cm}
\label{tab2} 
\begin{center} 
\begin{tabular}{c|cccccccccc} 
 \hline 
\multicolumn{11}{c}{ $J_i^\pi = 0^+_1,T=1$} \\
\hline
       $n$     & 0    & \multicolumn{3}{c}{2}    & \multicolumn{4}{c}{4}      & \multicolumn{2}{c}{6}      \\ 
\hline
 $A_{n}(0)$  & 0.33 & \multicolumn{3}{c}{0.34} & \multicolumn{4}{c}{0.13}   & \multicolumn{2}{c}{0.02}   \\  
\hline 
 $A_{n}(1)$  & 0.0 & \multicolumn{3}{c}{0.09} & \multicolumn{4}{c}{0.07}   & \multicolumn{2}{c}{0.01}   \\  
\hline 
\multicolumn{11}{c}{ $J_i^\pi = 1^+,T=1$ ($E_x=12.96$ MeV)} \\
\hline
      $n$     & 0    & \multicolumn{3}{c}{2}    & \multicolumn{4}{c}{4}      & \multicolumn{2}{c}{6}      \\
\hline 
 $A_{n}(0)$  & 0.0 & \multicolumn{3}{c}{0.15} & \multicolumn{4}{c}{0.30}   & \multicolumn{2}{c}{0.09}   \\  
\hline 
 $A_{n}(1)$  & 0.05 & \multicolumn{3}{c}{0.13} & \multicolumn{4}{c}{0.19}   & \multicolumn{2}{c}{0.08}   \\  
\hline 
\multicolumn{11}{c}{ $J_i^\pi = 1^+,T=2$ ($E_x=18.22$ MeV) } \\
\hline
      $n$     & 0    & \multicolumn{3}{c}{2}    & \multicolumn{4}{c}{4}      & \multicolumn{2}{c}{6}      \\ 
\hline
 $A_{n}(0)$  & 0.0 & \multicolumn{3}{c}{0.04} & \multicolumn{4}{c}{0.27}   & \multicolumn{2}{c}{0.12}   \\  
\hline 
 $A_{n}(1)$  & 0.0 & \multicolumn{3}{c}{0.11} & \multicolumn{4}{c}{0.29}   & \multicolumn{2}{c}{0.16}   \\  
\hline 
\end{tabular} 
\end{center} 
\end{table*} 
We find that this state is dominated by $n$p$n$h cross-shell excitations with $n \leq 4$,
with 0p0h and 2p2h excitations amounting to about 75$\%$ of the wave function. 
Importantly, the pure $sd$-shell configuration contributes only
$33\%$ to the $^{38}$Ar ground state. The rest 
corresponds to excitations to the $pf$ shell and hence opens new 
channels for the fragmentation of the M1 strength distribution, to which we turn now.

We have calculated the M1 strength distributions by using the Lanczos
method with 200 iterations.  This is sufficient to achieve convergence
for the calculated energies of the lowest states. However, at higher
excitation energies the calculated M1 distributions represent only the 
total strength per energy interval rather than true states.
We note that we have scaled the spin part of the M1 operator
by a constant factor $q=0.75$, as it is customary in shell model
calculations. It has been shown that shell model calculations using 
such an effective
spin operator describe the M1 \cite{PNC98} and GT \cite{Martinez96,Lang95} transitions
of $pf$-nuclei quite well. 

\begin{figure}
\begin{center}
\includegraphics[scale=0.34]{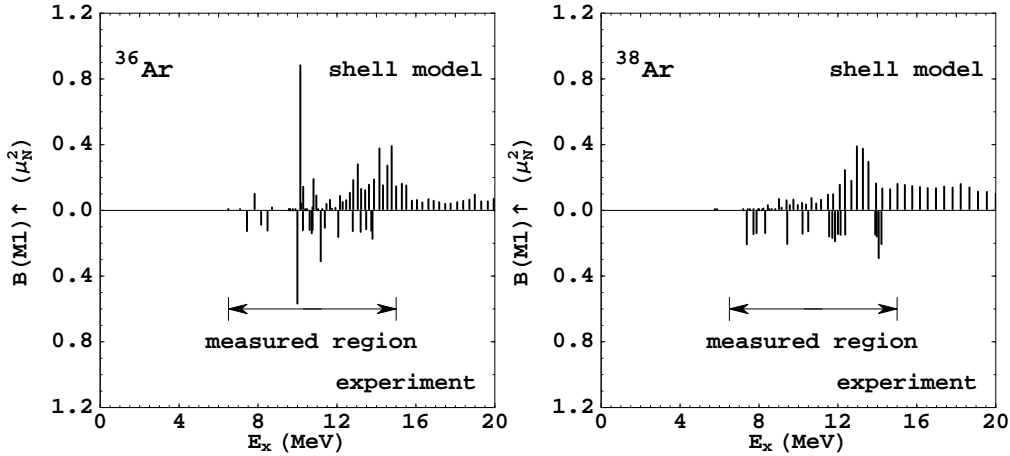}
\end{center}
\caption{Comparison of the experimental and theoretical M1 strength distributions 
for $^{36}$Ar (left) and $^{38}$Ar (right) calculated in the $sdp\!f_1$ model 
space.}
\label{fig6}
\end{figure}

The calculated  M1 strength distributions for $^{36}$Ar and $^{38}$Ar within the 
full $sdp\!f_1$ model space are shown in Fig.\ref{fig6}.
For $^{36}$Ar the result is similar
to the $sd$-shell calculation as, in agreement with
the experimental data, it shows a rather strong transition
at around 10 MeV. At higher energies,
the present $^{36}$Ar M1 strength distribution is significantly
more fragmented than the one obtained within the pure
$sd$-shell. Our calculation also predicts a noticeable amount of M1 strength 
outside of the experimental energy window. 
In fact, we calculate a total M1 strength for $^{36}$Ar of
6.76 $\mu_N^2$. Experimentally a summed M1 strength of
2.65(12) $\mu_N^2$ is observed up to an excitation energy
of 13.8 MeV, which is in agreement with the calculated value
(2.8 $\mu_N^2$). However, our calculation predicts an M1 strength of 
1.4 $\mu_N^2$ in the energy interval 13.8-15 MeV, where experimentally
no strength has been observed.

For $^{38}$Ar the result from the 
 $sdp\!f_1$ model space is strikingly different than that of the pure 
$sd$-shell calculation reported above. The calculated M1 strength in the energy
regime below 15 MeV is now strongly fragmented like the experimental one.
Moreover, we find a total M1 strength of 2.88 $\mu_N^2$ in the experimental 
energy window, to be compared with the experimental value 
of 2.86(18) $\mu_N^2$. For $^{38}$Ar (with a $T=1$ ground state) M1 
transition can lead to final states with $T=1$ and $T=2$. Most of 
the M1 strength to $T=2$ states resides at excitation energies above 15 MeV
so that the main peak of the M1 strength distribution around $E_x=13$ MeV
corresponds dominantly to $\Delta T=0$ transitions. 
As for $^{36}$Ar, our calculation predicts a strong part of the M1 strength
to reside at energies higher than the current experimental energy window.
In fact, our calculated total M1 strength is 6.01 $\mu_N^2$ for $^{38}$Ar.
It would be desirable to experimentally search for this predicted M1 strength
at higher energies.

To explore the origin of the fragmentation observed in the M1 distributions
for both argon isotopes, we have also analyzed the wave functions
of the strongest $1^+$ states.

\begin{figure}[t]
\begin{center}
\includegraphics[scale=0.34]{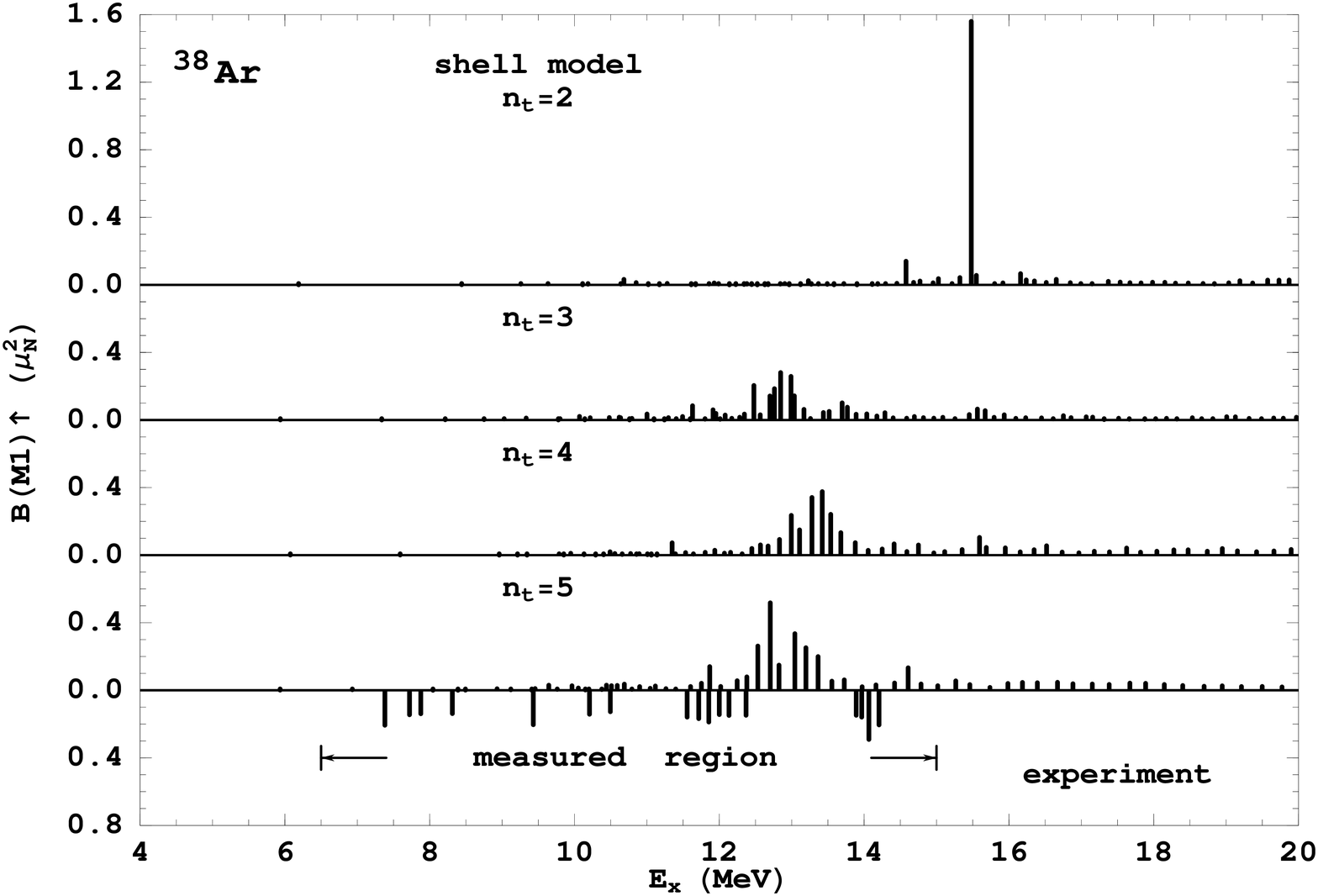}
\end{center}
\caption{Evolution of the M1 strength distribution with increase of the
model space. Here $n_t = n + n_{df}$, where $n$ is the number of particles 
excited across the $sd$--$pf$ gap and $n_{df} \leq 1$ stands for the 
number of nucleons promoted from the $d_{5/2}$ orbital to the $s_{1/2}d_{3/2}$ 
subspace and from the $f_{7/2}p_{3/2}$ to the $f_{5/2}p_{1/2}$ ones.} 
\label{fig7}
\end{figure}   
For $^{36}$Ar, Table \ref{tab1} lists the dominant configurations of 
the $1^+,T=1$ state at 10.14 MeV, which corresponds to the single 
strong M1 transition observed in the calculation (and in the data),
and the $1^+,T=1$ state at 14.77 MeV, which is a typical state in the
energy range of strongly fragmented M1 transitions. The $1^+$ state at 10.14 MeV is 
dominated by 0p0h and 2p2h configurations ($>80\%)$, similarly to the ground state.
We notice that the state has a quite significant 2p2h component 
($40\%$) with a simultaneous cross-shell excitation of
a proton and a neutron. 
Thus, we find, as in the pure $sd$-shell calculations, a strong M1 transition 
around 10 MeV, however, the configurations of the initial and final states are
quite different in the $sd$ and the $sdp\!f_1$ model spaces. 
The $1^+$ state at 14.77 MeV has sizable
4p4h configurations (about $30\%$), but in contrast it has a rather small
0p0h component (less than $10\%$). In this energy range the 
various configurations are quite strongly mixed resulting in the strong 
fragmentation of the M1 strength.

For $^{38}$Ar Table \ref{tab2} lists the dominant configurations of the
$1^+,T=1$ state at 12.96 MeV which carries a sizable
fraction of the fragmented M1 strength below 15 MeV. Different to the ground 
state, 0p0h excitations (pure $sd$-shell configurations) amount only to 
$5\%$, while 2p2h and 4p4h excitations are dominant. But importantly we also find
a noticeable contribution of 6p6h configurations in these states.
It is further worth mentioning that neutron excitations are more
important than proton excitations.

Obviously there are many $1^+$ states with large configuration mixing at moderate 
excitation energies which have a substantial $pf$-shell contribution, but 
only a rather small pure $sd$-shell component or a nearly vanishing
0p0h component. Table 2 lists the dominant configurations
of the $1^+,T=2$ state at 18.22 MeV, which is a typical
example in this energy regime.

\begin{figure}
\begin{center}
\includegraphics[scale=0.34]{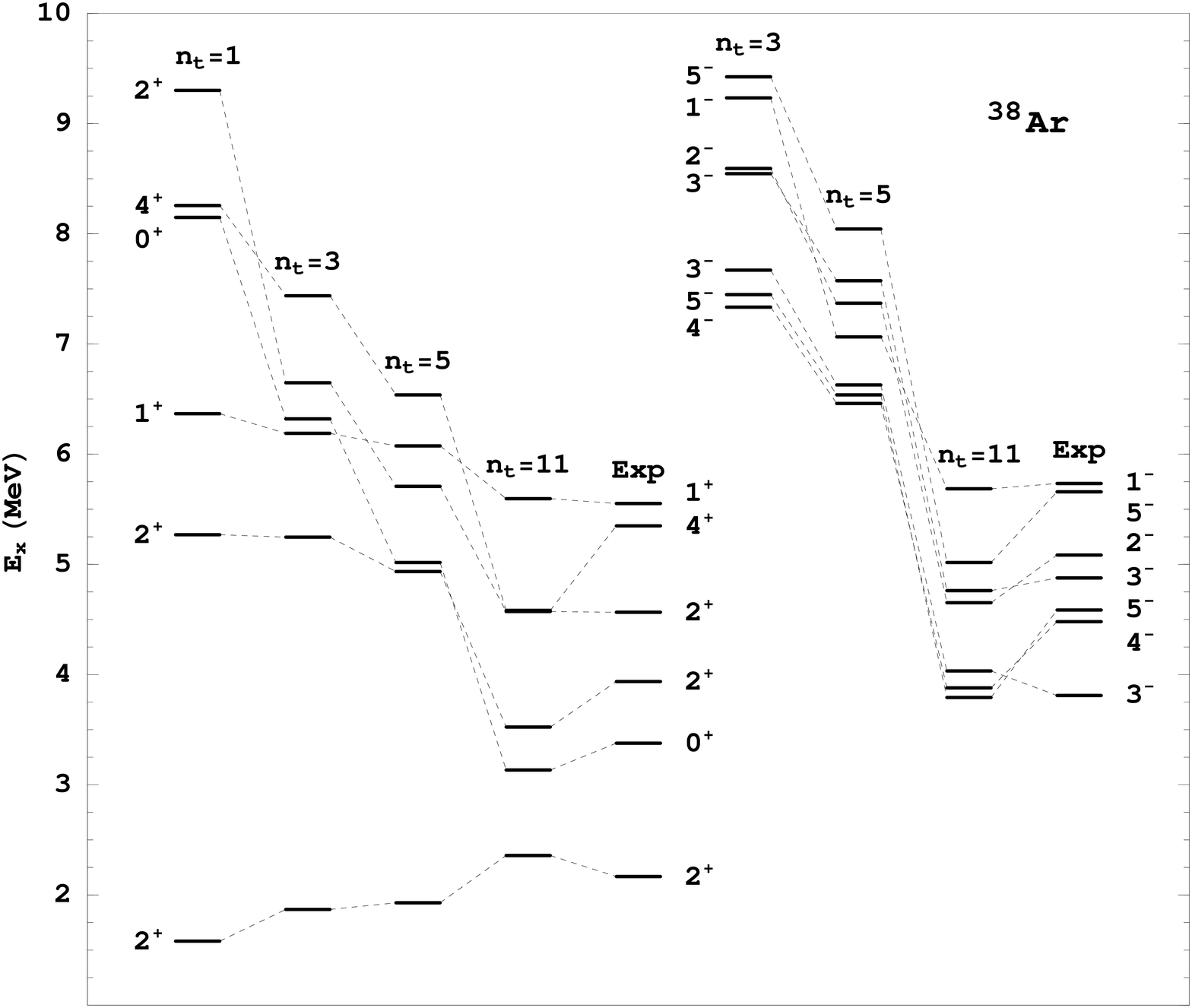}
\end{center}
\caption{Comparison of experimental and calculated spectra for $^{38}$Ar
in the $sdp\!f_1$ space for different values of $n_t=n+n_{df}$ with 
$n_{df}$ fixed at 1 and varying $n$ corresponding to $n$p$n$h cross-shell 
excitations. The full $sdp\!f_1$ space calculation corresponds to $n_t=11$.}
\label{fig8}
\end{figure}

Finally we have explored the evolution of the M1 strength distribution in $^{38}$Ar 
with the number of allowed cross-shell excitations.  It is convenient to 
characterize the different truncations using the quantity $n_t=n+n_{df}$, 
where $n$ is the number of particles moved across the $sd$--$pf$ shell gap 
and $n_{df}$ indicates the number of nucleons moved from the $d_{5/2}$ orbital to 
the $s_{1/2}d_{3/2}$ subspace and from the $f_{7/2}p_{3/2}$ subspace to 
the spin-orbit partners $f_{5/2}p_{1/2}$. In our $sdp\!f_1$ calculations we have 
allowed only 1p1h excitations between spin-orbit partners, hence   
$n_{df}\leq 1$.
The simplest excitation scheme is $n_t=2$ which allows the mixing of 
three configurations, namely 
the one with $n=0,n_{df}=0$, $n=2,n_{df}=0$ and  $n=0,n_{df}=1$. 
(Note that $n=1$ excitations are not possible due to parity.) The $n_t=2$ 
excitations do not change qualitatively the 
shape of the M1 strength function (see Fig. \ref{fig7}) 
as compared to the pure $sd$-shell results. Only the centroid is moved  
to higher energies, as the original single spin-flip state is shifted up 
in energy and is 
connected by the M1 operator only to the $n=0,n_{df}=0$ configuration. 
If the number of excitations is increased to 
$n_t=3$, only one new configuration with $n=2,n_{df}=1$ is added.
This configuration is, however,  
crucial for the fragmentation of the M1 strength (see Fig. \ref{fig7}). 
Besides the large M1 matrix element between the $n=0,n_{df}=0$ and 
$n=0,n_{df}=1$ configurations in the $sd$-shell, there is now another big 
M1 transition in the model space (between the $n=2,n_{df}=0$ and 
$n=2,n_{df}=1$ configurations) involving 2p2h excitations across the shell gap.
 The interference of these two contributions produces the fragmentation of the 
M1 strength. The further addition of new components like $n=4,n_{df}=0$ ($n_t=4$) 
and  $n=4,n_{df}=1$ ($n_t=5$) does not change the fragmentation picture qualitatively, 
however, the summed B(M1) strength increases from 2.95 $\mu_N^2$ ($n_t=3$ case) 
to  4.02 $\mu_N^2$ and 4.20 $\mu_N^2$, respectively,
as more $pf$-shell components are mixed into the ground
state wave function.  We also observe that the convergence 
for the energies of many states is not even achieved  at the $n_t=5$ level 
(see Fig. \ref{fig8}).  This indicates that more complicated configurations with 
$n_t>5$  have to be included, which we have done in our complete $n=10,n_{df}=1$
($n_t=11$) calculation discussed above.

\section{Conclusion}

We have performed large-scale shell model calculations for 
$^{36}$Ar and $^{38}$Ar within the complete $s_{1/2}d_{3/2}f_{7/2}p_{3/2}$ model 
space and additionally allowing 1p1h excitations from $d_{5/2}$ 
to the $s_{1/2}$, $d_{3/2}$ orbitals or from $f_{7/2}$ and $p_{3/2}$
to the rest of the $pf$-shell.
We reproduce the energy spectra for both isotopes,
including the negative parity states. Importantly our calculations
identify sizable contributions of cross-shell components in the 
$^{36}$Ar and $^{38}$Ar ground states, implying the onset of erosion 
of the N=20 shell closure. However, this erosion 
is a slow process involving many $n$p$n$h excitations until, for example, convergence
of the energies of the lowest excited states is achieved.
These studies reveal the onset of fragmentation in the M1 strength distributions.
However, only a marginal agreement with the data has been obtained, which might
be improved if higher-order cross-shell excitations are considered.

These cross-shell excitations play an important role for the M1 strength
distributions. For $^{36}$Ar our calculation exhibits a single 
strong M1 transition at 10 MeV, which is in agreement with the measured 
distribution. At higher excitation energies the calculated distribution 
is quite strongly fragmented, in contrast to the results of pure $sd$-shell 
calculations. Most of this fragmented strength resides, however, outside of 
the energy range for which currently data exist. 

For $^{38}$Ar our calculated M1 distribution is in even stronger contrast to the one
obtained in the pure $sd$-shell. The cross-shell $n$p$n$h excitations lead to
a very strong fragmentation of the strength. Particularly important are here
$(s_{1/2}d_{3/2})^{10-n}(f_{7/2}p_{3/2})^n$ configurations 
with $n=2,4$ and 6 which are strongly mixed in the 
wave functions of $1^+$ states above 10 MeV. Consequently, also 
the configurations which include 1p1h spin-flip excitations involving the 
$d_{5/2}$, $f_{5/2}$ and $p_{1/2}$ partners build on top of the $n$p$n$h
cross-shell excitations get strongly mixed as well.
As a consequence the M1 strength distribution is strongly fragmented,
as observed in the data. As for $^{36}$Ar our calculation
predicts also for $^{38}$Ar a sizable amount of M1 strength to reside 
at energies higher than the current observational limit.  

 Very recent 
experimental data  \cite{Chu06} obtained via the $^{40}$Ar($\vec{\gamma},\gamma'$) photon
scattering reaction suggest a very strong fragmentation of the M1
strength in $^{40}$Ar, where only one $1^+$ state with a B(M1) value of 0.145(59) $\mu_N^2$
has been identified in the energy region between 7.7 and 11 MeV. 
The evolution of the M1 strength with  increase of the neutron 
number ($N > 20$) for Ar isotopes is an intriguing issue which will contribute further to our
understanding of the $N=20$ cross-shell dynamics.

\end{document}